\documentclass[longauth]{aa}  
\usepackage{graphicx}
\usepackage{txfonts}
\usepackage{amsmath}	
\usepackage{hyperref}
\hypersetup{colorlinks=true, citecolor=blue}

\newcommand{\HII}{\ion{H}{II}}
\newcommand{\Ha}{H{$\alpha$}}
\newcommand{\kms}{\rm km\,s$^{-1}$}
\def \kms {{\rm km\,s$^{-1}$}}

\def \sol {{\rm M$_\odot$}}
\def \Lsol {{\rm L$_\odot$}}

\def \Kcmcb {{\rm K\,cm$^{-3}$}}

\def \Lha {\ensuremath{L_{\mathrm{H}\alpha}}}
\def \Lbol {\ensuremath{L_\mathrm{bol}}}

\def \Te {\ensuremath{T_\mathrm{e}}}
\def \ne {\ensuremath{n_\mathrm{e}}}

\def \pdir {\ensuremath{P_\mathrm{rad}}}
\def \ptherm {\ensuremath{P_\mathrm{therm}}}
\def \pwind {\ensuremath{P_\mathrm{wind}}}

\begin{document}

   \title{Linking stellar populations to HII regions across nearby galaxies}
   \subtitle{I. Constraining  pre-supernova feedback from young clusters in NGC1672}

\newcommand{\aifa}{Argelander-Institut f\"{u}r Astronomie, Universit\"{a}t Bonn, Auf dem H\"{u}gel 71, 53121, Bonn, Germany}
\newcommand{\ita}{Institut f\"{u}r Theoretische Astrophysik, Zentrum f\"{u}r Astronomie der Universit\"{a}t Heidelberg, Albert-Ueberle-Str 2, D-69120 Heidelberg, Germany}
\newcommand{\ari}{Astronomisches Rechen-Institut, Zentrum f\"{u}r Astronomie der Universit\"{a}t Heidelberg, M\"{o}nchhofstra\ss e 12-14, 69120 Heidelberg, Germany}
\newcommand{\princeton}{Department of Astrophysical Sciences, Princeton University, Princeton, NJ 08544 USA}
\newcommand{\arcetri}{INAF — Osservatorio Astrofisico di Arcetri, Largo E. Fermi 5, I-50125, Florence, Italy}
\newcommand{\mpe}{Max-Planck-Institut f{\"u}r extraterrestrische Physik, Giessenbachstra{\ss}e~1, D-85748 Garching, Germany}
\newcommand{\zah}{Universit\"{a}t Heidelberg, Zentrum f\"{u}r Astronomie, Albert-Ueberle-Str. 2, 69120 Heidelberg, Germany}
\newcommand{\eso}{European Southern Observatory, Karl-Schwarzschild-Stra{\ss}e 2, 85748 Garching, Germany}
\newcommand{\ljmu}{Astrophysics Research Institute, Liverpool John Moores University, 146 Brownlow Hill, Liverpool L3 5RF, UK}
\newcommand{\stern}{Sternberg Astronomical Institute, Lomonosov Moscow State University, Universitetsky pr. 13, 119234 Moscow, Russia}
\newcommand{\alberta}{Department of Physics, University of Alberta, Edmonton, AB T6G 2E1, Canada}
\newcommand{\ohio}{Department of Astronomy, The Ohio State University, 140 West 18th Avenue, Columbus, OH 43210, USA}
\newcommand{\wyoming}{Department of Physics \& Astronomy, University of Wyoming, Laramie, WY 82071}
\newcommand{\mpia}{Max-Planck-Institute for Astronomy, K\"onigstuhl 17, D-69117 Heidelberg, Germany}
\newcommand{\iwr}{Universit\"{a}t Heidelberg, Interdisziplin\"{a}res Zentrum f\"{u}r Wissenschaftliches Rechnen, 69120 Heidelberg, Germany}
\newcommand{\anu}{Research School of Astronomy and Astrophysics, Australian National University, Weston Creek, ACT 2611, Australia}
\newcommand{\Nihon}{Department of Physics, General Studies, College of Engineering, Nihon University, 1 Nakagawara, Tokusada, Tamuramachi, Koriyama, Fukushima, 963-8642, Japan} 
\newcommand{\naoj}{National Astronomical Observatory of Japan, 2-21-1 Osawa, Mitaka, Tokyo, 181-8588, Japan} 
\newcommand{\usyd}{Sydney Institute for Astronomy, School of Physics, Physics Road, The University of Sydney, Darlington 2006, NSW, Australia} 
\newcommand{\UGent}{Sterrenkundig Observatorium, Universiteit Gent, Krijgslaan 281 S9, B-9000 Gent, Belgium}
\newcommand{\carnegie}{The Observatories of the Carnegie Institution for Science, 813 Santa Barbara Street, Pasadena, CA 91101, USA} 
\newcommand{\Uchile}{Departamento de Astronom\'ia, Universidad de Chile, Camino del Observatorio 1515, Las Condes, Santiago, Chile}
\newcommand{\nrao}{National Radio Astronomy Observatory, 520 Edgemont Road, Charlottesville, VA 22903, USA}
\newcommand{\UWaus}{International Centre for Radio Astronomy Research, University of Western Australia, 7 Fairway, Crawley, 6009, WA, Australia}
\newcommand{\UTol}{Ritter Astrophysical Research Center, The University of Toledo, Toledo, OH 43606, USA}
\newcommand{\CITEVA}{Centro de Astronom\'ia (CITEVA), Universidad de Antofagasta, Avenida Angamos 601, Antofagasta, Chile}
\newcommand{\UWyom}{Department of Physics \& Astronomy, University of Wyoming, Laramie, WY 82071 USA}
\newcommand{\leon}{Univ Lyon, ENS de Lyon, CNRS, Centre de Recherche Astrophysique de Lyon UMR5574, F-69230 Saint-Genis-Laval France}
\newcommand{\Gemini}{Gemini Observatory/NSF’s NOIRLab, 950 N. Cherry Avenue, Tucson, AZ, 85719, USA}
\newcommand{\Steward}{Steward Observatory, University of Arizona, 933 North Cherry Avenue, Tucson, AZ 85721}
\newcommand{\UTamkang}{Department of Physics, Tamkang University, No.151, Yingzhuan Road, Tamsui District, New Taipei City 251301, Taiwan}
\newcommand{\Umadrid}{Departamento de F\'isica Te\'orica, Universidad Autónoma de Madrid, 28049 Cantoblanco, Spain}
\newcommand{\stsci}{Space Telescope Science Institute, Baltimore, MD 21218, USA}
\newcommand{\Ucal}{Department of Physics \& Astronomy, University of California, Riverside, CA, USA}
\newcommand{\TAPIR}{TAPIR, California Institute of Technology, Pasadena, CA 91125, USA}
\newcommand{\UDurEA}{Centre for Extragalactic Astronomy, Department of Physics, Durham University, South Road, Durham DH1 3LE, UK}
\newcommand{\UDurCC}{Institute for Computational Cosmology, Department of Physics, University of Durham, South Road, Durham DH1 3LE, UK}
\newcommand{\JHop}{Department of Physics \& Astronomy, Bloomberg Center for Physics and Astronomy, Johns Hopkins University, 3400 N. Charles Street, Baltimore, MD 21218}

\authorrunning{A.~T.~Barnes et al. }

\author{A.~T.~Barnes,\inst{1}\thanks{\url{ashleybarnes.astro@gmail.com}}
        R.~Chandar,\inst{2}
        K.~Kreckel,\inst{3}
        S.~C.~O.~Glover,\inst{4}
        F.~Scheuermann,\inst{3}
        F.~Belfiore,\inst{5}
        F.~Bigiel,\inst{1}
        G.~A.~Blanc,\inst{6,7}
        M.~Boquien,\inst{8}
        J.~den Brok,\inst{1}
        E.~Congiu,\inst{7}
        M.~Chevance,\inst{3}
        D.~A.~Dale,\inst{9}
        S.~Deger,\inst{10}
        J.~M.~D.~Kruijssen,\inst{3}
        O.~V.~Egorov,\inst{3}
        C.~Eibensteiner,\inst{1}
        E.~Emsellem,\inst{11,12}
        K.~Grasha,\inst{13}   
        B.~Groves,\inst{14} 
        R.~S.~Klessen,\inst{4,15} 
        S.~Hannon,\inst{16} 
        H.~Hassani,\inst{17} 
        J.~C.~Lee,\inst{18,19} 
        A.~K.~Leroy,\inst{20} 
        L.~A.~Lopez,\inst{20} 
        A.~F.~McLeod,\inst{21,22} 
        H.~Pan,\inst{23} 
        P.~S\'anchez-Bl\'azquez,\inst{24} 
        E.~Schinnerer,\inst{25} 
        M.~C.~Sormani,\inst{4}
        D.~A.~Thilker,\inst{26}
        L.~Ubeda,\inst{27} 
        E.~J.~Watkins,\inst{3}
        T.~G.~Williams\inst{25} 
        }

\institute{\aifa \and 
        \UTol \and
        \ari \and
        \ita \and 
        \arcetri \and
        \carnegie \and
        \Uchile \and
        \CITEVA \and
        \UWyom \and
        \TAPIR \and
        \eso \and
        \leon \and
        \anu \and
        \UWaus \and
        \iwr \and
        \Ucal \and
        \alberta \and
        \Gemini \and
        \Steward \and
        \ohio \and
        \UDurEA \and
        \UDurCC \and
        \UTamkang \and
        \Umadrid \and
        \mpia \and
        \JHop \and
        \stsci}

   \date{Received -; accepted -}

 
  \abstract
  {One of the fundamental factors regulating the evolution of galaxies is {\it stellar feedback}.
  However, we still do not have strong observational constraints on the relative importance of the different feedback mechanisms (e.g.\ radiation, ionised gas pressure, stellar winds) in driving \HII\ region evolution and molecular cloud disruption. To quantify and compare the different feedback mechanisms, the size of an \HII\ region is crucial, but samples of well-resolved \HII\ regions are scarce.}
   {We constrain the relative importance of the various feedback mechanisms from young massive star populations by resolving \HII\ regions across the disk of the nearby star-forming galaxy NGC~1672.}
   {We combine measurements of ionised gas nebular lines obtained by PHANGS-MUSE, with high-resolution (PSF FWHM$\sim$\,0.1\arcsec; $\sim$\,10 pc) imaging from the {\em Hubble Space Telescope} (HST) in both the narrow-band H$\alpha$ and broad-band (NUV, {\it U}, {\it B}, {\it V}, {\it I}) filters.
   We identify a sample of 40 isolated, compact \HII\ regions in the HST H$\alpha$ image. 
   We measure the sizes of these \HII\ regions, which were previously unresolved in seeing-limited ground-based observations. 
   In addition, we identify the ionisation source(s) for each \HII\ region from catalogues produced as part of the PHANGS-HST survey.
   In doing so, we are able to connect young stellar populations with the properties of their surrounding \HII\ regions.}
   {
   The HST observations allow us to resolve all 40 regions, which have radii between 5 to 40\,pc.
   The \HII\ regions investigated are mildly dominated by thermal or wind pressure, yet their elevation above the radiation pressure is within the expected uncertainty range.
   We see that radiation pressure provides a substantially higher contribution to the total pressure than previously found in the literature over similar size scales.
   In general, we find higher pressures within more compact \HII\ regions, which is driven by the inherent size scaling relations of each pressure term, albeit with significant scatter introduced by the variation in the stellar population properties (e.g.\ luminosity, mass, age, metallicity).
   }
    {For nearby galaxies, the combination of MUSE/VLT observations with stellar population and resolved H$\alpha$ observations from HST provides a promising approach that could yield the statistics required to map out how the importance of different stellar feedback mechanisms evolve over the lifetime of an \HII\ region.} 
   \keywords{ (ISM:) HII regions -- Galaxies: evolution -- Galaxies: star clusters: general}

   \maketitle
%

\section{Introduction}

High-mass stars (${>}8$\,\sol) are fundamental for driving the evolution of galaxies, due to the large amounts of energy and momentum (i.e.\ stellar feedback) that they inject into the interstellar medium (ISM) during their short lifetimes \citep[e.g.][]{krumholz_2014}.
Recent simulations \citep[e.g.][]{dale_2012, dale_2013, Raskutti2016, Gatto2017, Rahner2017, Rahner2019, Kim2018, Kim_JG2021, Kannan2020, Jeffreson2021} and observational evidence \citep[e.g.][]{Grasha2018, Grasha2019, kruijssen19a, chevance20b, chevance20, Kim2021,McLeod2021,Barrera-Ballesteros2021a,Barrera-Ballesteros2021b, Hannon2019, Hannon2022} suggest that, in particular, feedback in the early (pre-SNe) stages of high-mass stars (i.e. within \HII\ regions) plays the critical role in driving the evolution of both its local and larger scale environment. 
Hence, there has been considerable interest in observationally quantifying the impact of the various early (pre-SN) feedback mechanisms in \HII\ regions  \citep[e.g.][]{pellegrini_2011,lopez_2014,mcleod_2019,McLeod2021,Kruijssen2019,Chevance2022a,Chevance2022b,olivier2020,barnes20b,Barnes2021b}. 


\begin{figure*}
    \centering
	\includegraphics[width=\textwidth]{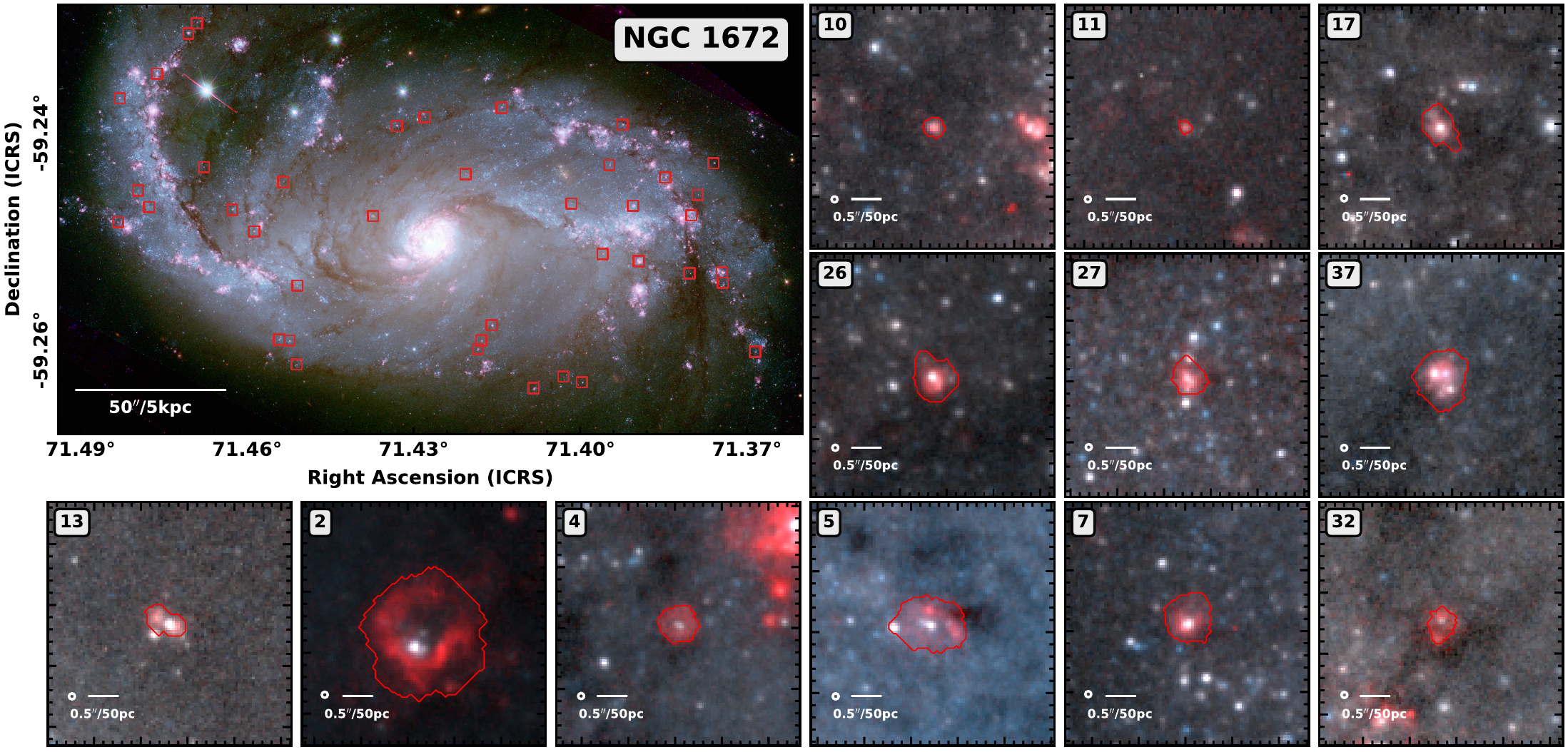}
    \caption{{\bf \HII\ regions identified towards NGC\,1672.} The three colour images are composed of 814\,nm, 555\,nm and 435\,nm wideband filters from the HST \citep{Jenkins2011,Lee2021}, and in red the HST continuum-subtracted 658\,nm (\Ha) narrow-band emission \citep{Jenkins2011}. 
    ({\it upper left}) Map of the galaxy disc overlaid with the boxes showing the  positions of the \HII\ region sample. 
    ({\it right and below}) Image cutouts for a subset of the \HII\ regions. The region ID is shown in the upper left of each panel (Tab.\,\ref{tab1}). Overlaid as a red contour is the boundary of each \HII\ region (\S\,\ref{sec:HIIregionProps}). In the lower left of each panel we show a scale bar, and a circle denoting the approximate HST PSF size (0.1\arcsec, 9\,pc).}
    \label{fig:rgb_main}
\end{figure*}

The dynamics and expansion of \HII\ regions may be driven by several possible sources of internal energy and momentum injection. 
A useful quantity to observationally constrain and compare these different feedback mechanisms is the pressure.
For example, \HII\ regions are filled with warm (\Te\,$\sim$\,$10^{4}$\,K) ionised hydrogen, which imparts an outward gas pressure \citep[e.g.][]{spitzer_1978}.
This thermal pressure of this photoionised gas, \ptherm, is set by the ideal gas law, 
\begin{equation}
\ptherm/k_\mathrm{B} = (\ne+n_\mathrm{H}+n_\mathrm{He}) \Te\ \approx 2\ne \Te~, 
\label{eq:ptherm}
\end{equation}
where $\ne$, $n_\mathrm{H}$, and $n_\mathrm{He}$ are the electron, H and He number densities, respectively, \Te\ is the electron temperature, and $k_\mathrm{B}$ is the Boltzmann constant.
Here, all He is assumed to be singly ionised. 
Constraints on the electron density, \ne, can be obtained from e.g. a Strömgren sphere approximation with accurate measurements of the source radius, $r$, and ionisation rate, $Q$ (\ne\,$\propto$\,$r^{3/2}$; \citealp{stromgren_1939}).     

An additional source of feedback is the intense radiation field produced by young, massive stellar populations. 
Assuming the stellar population's bolometric luminosity (\Lbol) emerges in the UV or optical and photons are absorbed once at the \HII\ region inner shell (for alternative see e.g.\ \citealp{Krumholz2009,draine_2011,Reissl2018}), the volume-averaged direct radiation pressure is given as \citep[e.g.][]{lopez_2014},
\begin{equation}
\pdir/k_\mathrm{B} = \frac{3 \Lbol}{4 \pi r^{2} c k_\mathrm{B}}~,
\label{eq:pdir}
\end{equation}
where $c$ is the speed of light.

Lastly, in their early evolutionary stages, high-mass stars produce strong stellar winds that can result in mechanical pressure within \HII\ regions. 
The wind ram pressure is calculated as,
\begin{equation}
\pwind/k_\mathrm{B} = \frac{3 \dot M v_\mathrm{wind}}{4\pi r^{2}k_\mathrm{B}}~,
\label{eq:pwind}
\end{equation}
where $\dot M$ is the mass loss rate and $v_\mathrm{wind}$ is the wind velocity (\S\,\ref{sec:prescalc}).


Using the above, recently \citet{Barnes2021b} assessed the magnitudes of feedback mechanisms acting within a sample of $\sim$\,6000 \HII\ regions identified from the PHANGS-MUSE survey of 19 nearby ($<$\,20\,Mpc)  star-forming, main sequence spiral galaxies \citep{Emsellem2022}. 
However, in this work, the majority of the \HII\ regions remained unresolved by the ${\sim}50{-}100$\,pc resolution of the ground-based observations, only allowing for limits to be placed on the size of these \HII\ regions. 
These large uncertainties on size measurements, and missing knowledge of the detailed morphology (such as broken shells, suggestive of a large escape fraction of ionising gas and winds), result in a large uncertainty (of about two orders of magnitude) in their pressure calculations.  
In this work, we directly address this uncertainty by introducing new size measurements from high-resolution HST data --- {\it key for constraining the above pressure terms} --- for a subsample of isolated, compact \HII\ regions (see Fig.\,\ref{fig:rgb_main}).

\section{Observations of NGC~1672}


We make use of HST and VLT/MUSE observations towards a single galaxy: NGC~1672 (see \citealp{Emsellem2022, Lee2021} for survey overviews).  
NGC 1672 is a nearby (19.4\,$\pm$\,2.9\,Mpc; \citealp{Anand2021}), strongly-barred and actively star-forming spiral galaxy. 
It is a good candidate for this initial study thanks to its high star formation rate (7.6\,$\mathrm{M_{\odot}\,yr^{-1}}$; \citealp{Leroy2021a}), which yields a large sample of \HII\ regions to study, and moderate inclination ($i$\,$\sim$\, 40$^{\circ}$; \citealp{Lang2020}), which limits the effects of extinction and line-of-sight confusion.  

\subsection{HST}


We make use of HST observations from \citet{Jenkins2011} and the PHANGS-HST survey \citep{Lee2021}. The PSF of these observations has a FWHM of $\sim$\,0.07\,$-$\,0.1\arcsec\ (7\,$-$9\,pc) depending on the filter, and the field of view (FOV) covers the majority of the galaxy disc. 

\medskip
\noindent{\bf Broad-band Observations:} We use observations made with the F435W and F814W ({\it I}) filters on the Advanced Camera for Surveys (ACS) taken from project 10354 \citep{Jenkins2011}, and with the F275W (NUV), F336W ({\it U}) and F555W ({\it V}) filters using the Wide Field Camera 3 (WFC3; UVIS) taken as part of the PHANGS-HST treasury
program 15654 (\citealp{Lee2021}).
The PHANGS-HST UVIS and archival ACS data were both reduced as part of the PHANGS survey (see \citealp{Lee2021}).


\medskip
\noindent{\bf Narrow/medium-band Observations:} We also include archival ACS observations using the F658N (i.e including \Ha) and F550M filters taken as part of project 10354. 
The narrow-band F658N map is continuum subtracted using an image formed from a combination of the F814W and F550M maps, appropriately scaled using their AB zero-points (see \citealp{Hannon2022} for method).

\subsection{PHANGS-MUSE}

We make use of VLT/MUSE observations from the PHANGS-MUSE survey (see \citealp{Emsellem2022} for a complete discussion of the processing and reduction of the MUSE observations). 
The PSF of these observations has a FWHM of 0.96\arcsec (90\,pc), and the FOV is comparable to the HST observations. 
The MUSE Integral Field Unit provides a typical spectral resolution (FWHM) of ${\sim}2.5$\,\AA\ (or ${\sim}100$\,\kms) covering lines (e.g. \Ha, H$\beta$, [\ion{S}{II}]) within the spectral range $4800{-}9300$\,\AA.
In this letter, we include properties (e.g. extinction corrected intensities) of individual \HII\ regions measured from their integral spectra from MUSE, as presented in ionised nebula catalogue from \citet{Santoro2022}. 


\section{Properties of \HII\ Regions and their Ionising Sources}

\subsection{Properties of \HII\ Regions}
\label{sec:HIIregionProps}

\begin{figure}
    \centering
	\includegraphics[width=\columnwidth]{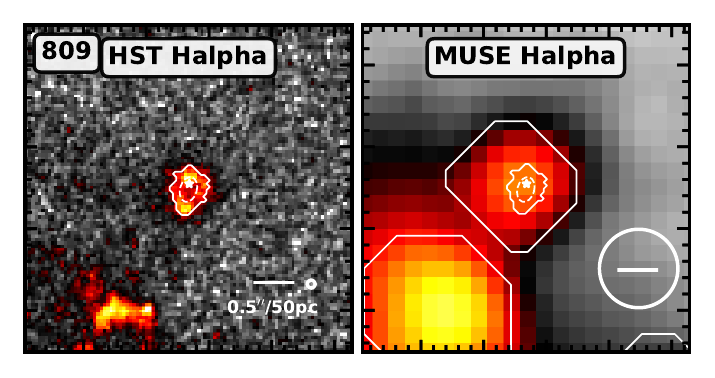}
    \caption{{\bf Comparison between the HST and MUSE observations towards one of the \HII\ regions in our sample}. 
    ({\em left}) The HST \Ha\ narrowband emission overlaid with a contour outlining the isophotal boundary as defined from our dendrogram analysis. 
    We show the intensity-weighted second order moment of the structure within this boundary as a dashed ellipse, which we use to define the radius of the \HII\ region (\S\,\ref{sec:HIIregionProps}). 
    A star marks the position of the identified ionising stellar population (\S\,\ref{sec:IonsingSourceProps}). 
    In the upper left is the ID from the MUSE nebula catalogue \citep{Santoro2022}.
    ({\em right}) The \Ha\ emission observed with MUSE \citep{Emsellem2022}, overlaid with white contours showing the boundaries of sources identified as part of the nebula catalogue from \citet{Santoro2022}. 
    In the lower right of each panel we show a scale bar, and a circle denoting the approximate PSF size.
    }
    \label{fig:ident}
\end{figure}

We use the high-resolution continuum subtracted HST \Ha\ map to resolve the structure of the \HII\ regions identified in the PHANGS-MUSE nebula catalogue (\citealp{Santoro2022}; see Fig.\,\ref{fig:ident}).   
Based on the following criteria, we select a sample of isolated 40 \HII\ regions out of the 1581 nebulae identified across NGC\,1672 in this catalogue (Fig.\,\ref{fig:rgb_main}). They:
\begin{itemize}
    \item[i)] meet the BPT and H$\alpha$ velocity dispersion criteria to be classified as a \HII\ region (e.g. see \citealp{Barnes2021b});
    \item[ii)] are sufficiently isolated such that they dominate the emission in MUSE line maps which are clearly, from manual inspection of the maps, uncontaminated by other sources;
    \item[iii)] contain a single \HII\ region with a compact morphology in the HST \Ha\ map, which has an apparent circular or simple shell-like profile in the image;
    \item[vi)] are spatially resolved (i.e. more extended than the PSF FWHM) in the HST \Ha\ map so that we can make a direct measurement of the radius.\footnote{We can robustly measure sizes that are fractions of a pixel broader than the PSF. For example, the Ishape software \citep{larsen99} can reliably measure the radii of compact sources with good $S/N$ that are only $\sim0.2$~pix broader than the PSF, which is less than 1\,pc at the distance of NGC\,1672.}
    \end{itemize}

Our regions appear in the HST H$\alpha$ images as discrete, well-defined sources.
To measure their sizes, we measure the noise in the H$\alpha$ images from a 5\arcsec\ cutout around each \HII\ MUSE region, and assign the contiguous region of pixels with $S/N > 5$ to represent the \HII\ region (see the red contours in Figs. \ref{fig:rgb_main} and \ref{fig:ident}). 
Then we measure their radius ($r$) by taking the geometric mean of the intensity-weighted spatial second moment using the H$\alpha$ intensity and assigned pixels (i.e. mean of the semi-major and semi-minor axis of ellipse shown in Fig.\,\ref{fig:ident}).\footnote{We carry out these calculations using the \textsc{astrodendro} software, though we do not use the full hierarchical information \citep[see][]{rosolowsky_2008}, only the outer contour.}
This radius approximately corresponds to the inner portion of the \HII\ region shell for extended sources, or one standard deviation around the peak of compact sources (see Fig.\,\ref{fig:ident}). 
This definition is physically motivated given that for \HII\ regions with shell-like morphologies, the centre is likely devoid of gas. 
To estimate the feedback process that is dominating their expansion, we compare the pressures at the inner edge of the shell (see \S~\ref{sec:prescalc}). 
Comparison between the measured size and flux radial profile for each \HII\ region suggests that, depending on geometry, the uncertainties in our radius measurements are no larger than 50\%.


\subsection{Properties of Ionising Sources} 
\label{sec:IonsingSourceProps}

We use the broad-band HST imaging to determine the stellar sources responsible for the ionisation of each \HII\ region.
These sources are typically close to the centre of each \HII\ region.  
We assume that even the most compact \HII\ regions in our sample are being ionised by a young stellar population rather than by an individual massive star.
This is justified as the \Ha\ luminosity completeness limit in the MUSE nebula catalogue for NGC~1672 is $\sim$\,$10^{37}$\,erg\,s$^{-1}$, which is higher than what can be produced by a single massive (e.g. O7V) star (see \citealp{Santoro2022}).

We follow the general procedure described in \citet{Turner21} to estimate the age and mass of each ionising stellar population, in other words, 
we use aperture photometry in each band performed through the PHANGS-HST pipeline using a 4-pixel radius (which is large enough to capture  most of the light from a cluster, yet small enough to avoid much contamination from nearby sources in crowded regions). 
The pipeline applies an  aperture correction to each filter determined from several isolated stellar populations, as described in \citet{Deger22}.
After correcting for the foreground extinction based on a Milky Way extinction law \citep{fitzpatrick_1999} and the foreground value given in NED, we fit the measured magnitudes with those predicted by the \citet{Bruzual2003} population synthesis models, assuming solar metallicity and a Chabrier stellar initial mass function (IMF; \citealp{chabrier_2003}). 
Predictions from the solar metallicity model provide a better overall fit than those from other metallicities to the measured broad-band colors of very young clusters (the ionizing sources of HII regions) in NGC~1672, consistent with the mean HII region value of 12+log(O/H)$\approx$8.56 measured from the MUSE spectra. The age estimates are mostly unchanged if we assume the $1/2 \times$ solar metallicity model instead.
We perform spectral energy distribution fitting using the publicly available CIGALE fitting package (e.g. \citealp{Boquien19}) to determine the best combination of age and local reddening (assuming a Galactic extinction law; \citealp{fitzpatrick_1999}) for each stellar population \citep{Turner21}.
The predicted mass-to-light ratio and extinction-corrected luminosity give an estimate of the stellar population mass.

We find the ionising sources all have estimated ages younger than 5~Myr (i.e. pre-SN feedback), and stellar masses in the range from 10$^{3}$ to 10$^{5}~\mathrm{M}_{\odot}$. 
Based on previous experiments, uncertainties in the age and mass of each young stellar population is approximately $\Delta$log$_{10}(t)\approx \Delta$log$_{10}(M) \approx 0.3$~dex, i.e.\ factor of $\sim$\,2 \citep[e.g.][]{Chandar10}.
In addition to these uncertainties, the lowest mass stellar populations within our sample ($<$~10$^{4}~\mathrm{M}_{\odot}$) may suffer from stochastic sampling effects of the IMF (e.g. \citealp{Krumholz2019}).
However, there is a limit to how deficient low mass clusters could be in high mass stars while still producing enough ionising radiation to produce a detectable \HII\ region, and all the \HII\ regions within our sample sit above this (bolometric luminosity) threshold; i.e.\ where stochastic effects are less significant ($L_{\rm bol} >$\,10$^{39}$\,erg\,s$^{-1}$; \citealp{daSilva2012}).




\section{Pressure calculations}\label{sec:prescalc}

\begin{figure}
    \centering
	\includegraphics[width=\columnwidth]{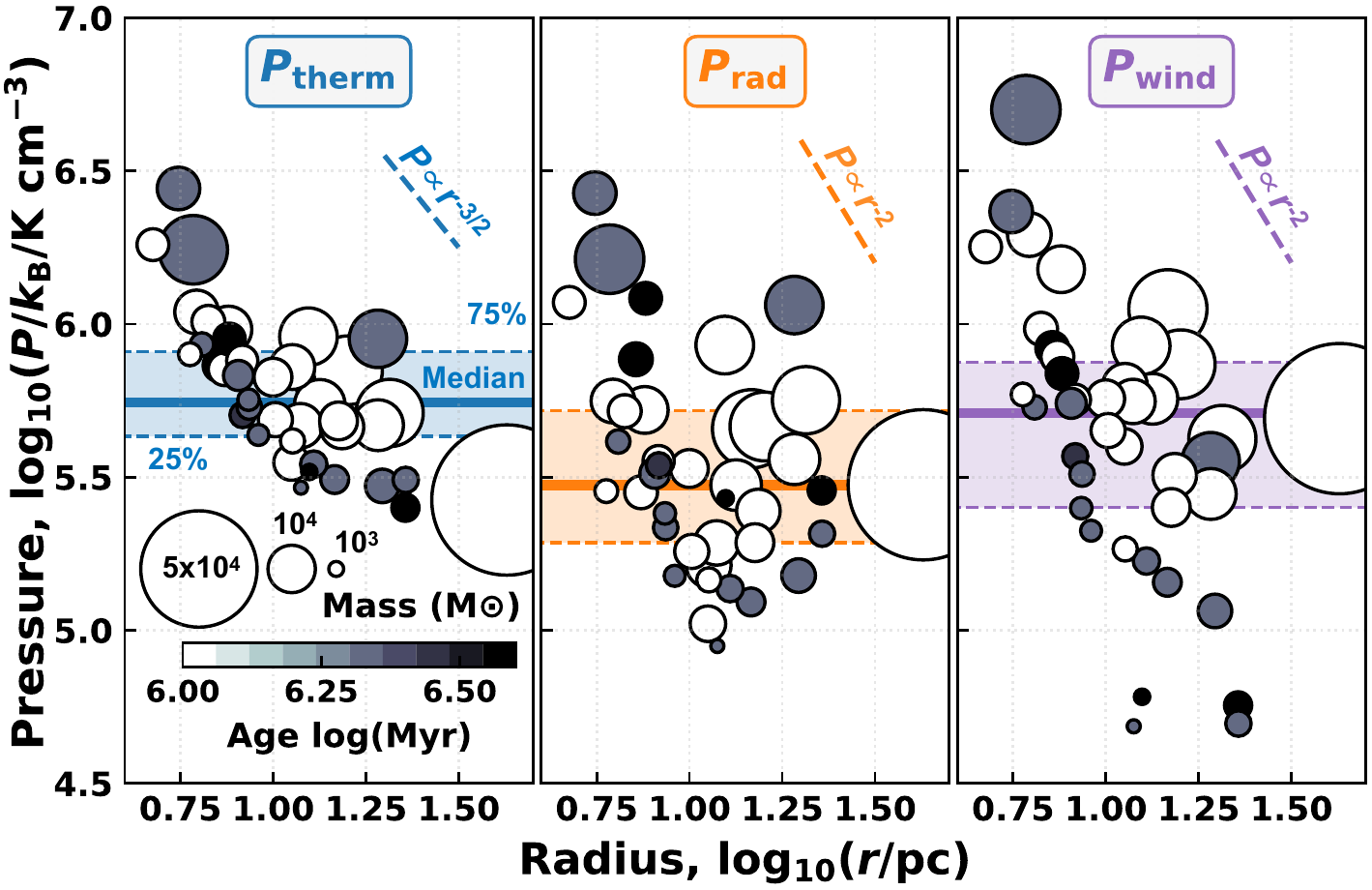}
    \caption{{\bf Distribution of the pressures due to the various feedback mechanisms as a function of \HII\ region radius, as well as stellar population mass (circle size) and age (circle colour).}
    We show a key for the size (i.e.\ mass) and a colour bar denoting the age at the bottom of the left panel. 
    The horizontal lines show the 25, 50 (median), and 75 percentiles for each pressure.
    The intrinsic relations of each pressure term are shown as a short dashed line in the upper right of each panel.}
    \label{fig:PvR}
\end{figure}



We determine \ptherm\ (Eq.\,\ref{eq:ptherm}) using values of the electron temperature (\Te) determined from the MUSE nitrogen auroral lines finding a range of 7000--11,000\,K or, for the 20 \HII\ regions without a significant \Te\ measurement (i.e.\ where the auroral \ion{N}{II} emission is too faint to be detected), we adopt a value of $\Te = 8000$\,K, corresponding approximately to the mean value of all \HII\ regions in NGC1672 with well-determined temperatures. 
We estimate the electron density assuming a smooth, spherical geometry as $\ne^2 = (3Q)/(4\pi r^{3} \alpha_\mathrm{B})$, where $Q$ is the ionisation rate (also determined using MUSE) and $\alpha_\mathrm{B}$ is the temperature-dependent case~B recombination coefficient, which is taken from \citet{Ferland92}. 
Here, $Q \approx (L_\mathrm{H\alpha} \alpha_\mathrm{B}) / (\alpha^\mathrm{eff}_\mathrm{H\alpha} h\nu_{H\alpha})$, where the effective recombination coefficient (i.e.\ the rate coefficient for recombinations resulting in the emission of an \Ha\ photon) is $\alpha^\mathrm{eff}_\mathrm{H\alpha} \approx 1.17\times10^{-13}$\,cm$^3$s$^{-1}$ \citep{Osterbrock2006}, $\nu_\mathrm{H\alpha}$ is the frequency of the \Ha\ emission line and $h$~is the Planck constant. 
For comparison, for the five sources for which we can derive a reliable estimate of \ne\ from the [\ion{S}{II}] emission line ratio, we find that \ne([\ion{S}{II}])>\ne($r$) by factors of $\sim$ two to three, which is expected given that the \HII\ regions show clumpy sub-structure (see discussion in \citealp{Barnes2021b}).
Given this bias, we use only our spherical geometry estimates of the electron density in our calculations of \ptherm.

To calculate \pdir\ (Eq.\,\ref{eq:pdir}), we compute \Lbol\ following \citet[][see their Fig. 7]{Barnes2021b}, where we 
used {\sc starburst99} models \citep{Leitherer1999} to determine a relation between the age of the stellar population ($t$) and \Lbol/\Lha, where \Lha\ is the extinction corrected \Ha\ luminosity from the MUSE observations. 
\begin{figure}
    \centering
	\includegraphics[width=\columnwidth]{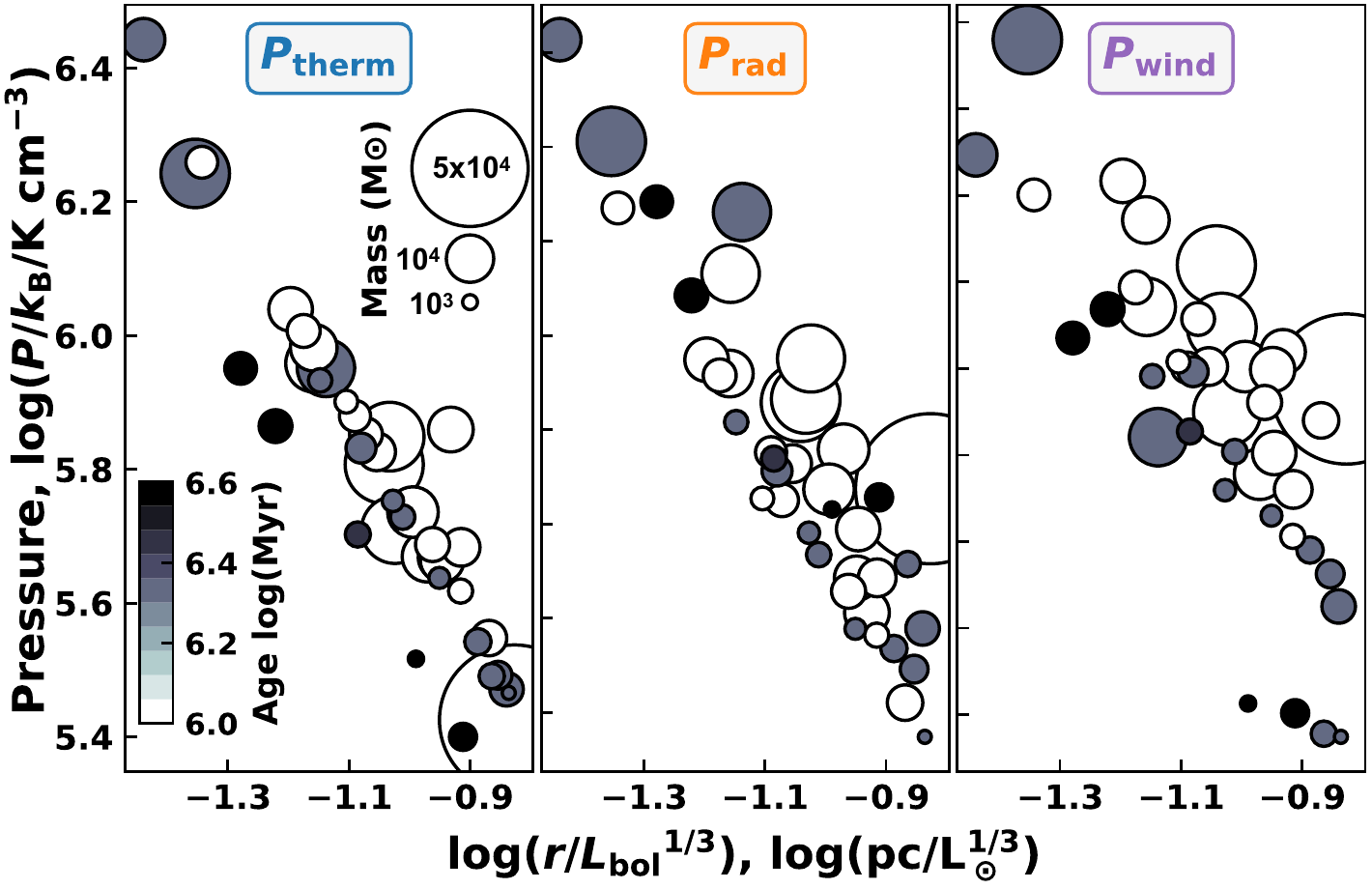}
    \caption{{\bf Pressures from the various feedback mechanisms as a function of radius normalised to the bolometric luminosity from the ionising stellar population.}
    We show a key for the size (i.e.\ mass) and a colour bar denoting the age in the left panel.}
    \label{fig:PvRL}
\end{figure}

To calculate \pwind\ (Eq.\,\ref{eq:pwind}), we estimate the wind velocity as, $v_\mathrm{wind} = ( 2 L_\mathrm{mech}/\dot M )^{0.5}$, where $L_\mathrm{mech}$ is the mechanical luminosity. 
Both $L_\mathrm{mech}$ and $\dot M$ taken from {\sc starburst99} models \citep{Leitherer1999}, where they vary as a function of the stellar population mass ($M$) and age ($t$) (see e.g. Fig. 8 \citealp{Barnes2021b}). To constrain $L_\mathrm{mech}$ and $\dot M$ and, hence, determine \pwind\ for each \HII\ region, we make use of the stellar population $M$ and $t$ determined from the HST broadband maps (\S\,\ref{sec:IonsingSourceProps}).

It is worth noting that Eqs.\,\ref{eq:ptherm}, \ref{eq:pdir} \& \ref{eq:pwind} intrinsically impose a radial dependence on all the pressure terms, and, in addition, both \ptherm\ and \pdir\ also depend on the \Ha\ luminosity. 
We discuss this in more detail when interpreting the correlated axes in the plots of the following section.

\section{Discussion and Summary}

\begin{figure}
    \centering
	\includegraphics[width=\columnwidth]{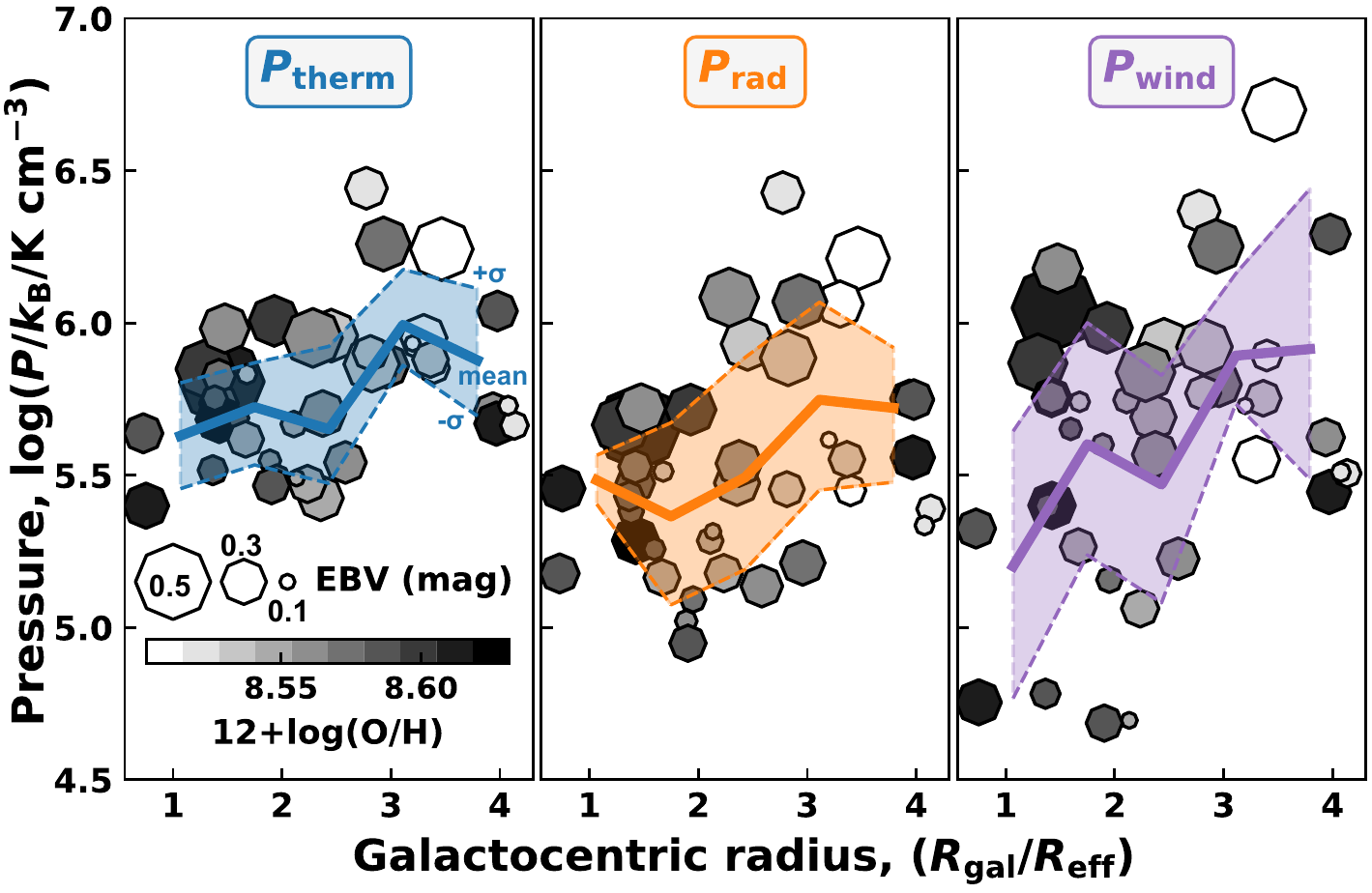}
    \caption{{\bf Distribution of the pressures due to the various feedback mechanisms as a function of galactocentric radius (normalised to the effective radius of 3.5\,kpc; \citealp{Leroy2021a}), as well as extinction (size) and metalicity (colour).}
    We show a key for the size (i.e.\ extinction) reference of each data point and a colour bar denoting the metalicity at the bottom of the left panel.
    The overlaid lines and shaded area show the mean and standard deviation of the data within equally spaced bins.}
    \label{fig:PvRmetal}
\end{figure}

We investigate how the different pressure terms vary as a function of the determined \HII\ region properties (Fig.\,\ref{fig:PvR}). 
We find that, in most cases, either thermal pressure (median $10^{5.74}$\,\Kcmcb) or wind ram pressure ($10^{5.71}$\,\Kcmcb) is mildly dominant. 
The radiation pressure is typically lower ($10^{5.47}$\,\Kcmcb), albeit within the factor of around two uncertainty on these measurements.
These results are in broad agreement with those in the literature made for \HII\ regions with sizes up to 10~pc (e.g.\ \citealp{lopez_2011,Rahner2017,mcleod_2019,mcleod_20}). 
Interestingly, however, the direct radiation pressure determined here for NGC\,1672 provides a substantially higher contribution to the total pressure than previous sources studied in the literature (e.g. LMC/SMC, NGC\,300; \citealp{lopez_2014, McLeod2021}). 
This could be a result of our sample bias toward luminous and compact \HII\ regions within this actively star-forming galaxy (7.6\,$\mathrm{M_{\odot}\,yr^{-1}}$; \citealp{Leroy2021a}), or differences in \HII\ region properties (e.g. metallicity). 

Fig.\,\ref{fig:PvR} also shows a trend of increasing pressure with decreasing radius. 
In other words, smaller \HII\ regions are typically more highly pressurised.
However, by definition (\S\,\ref{sec:prescalc}), the pressure terms are not independent of the radius (see intrinsic relations shown as dashed lines in Fig.\,\ref{fig:PvR}).
The significant observed scatter away from these simple power-law dependencies may then be introduced by the variation in the stellar population properties (e.g.\ luminosity).
We see that trends with the mass (i.e.\ circle size in Fig.\,\ref{fig:PvR}) or age (circle colour) of the ionising stellar population responsible for the \HII\ regions are, however, less clear than with size.
There is tentative evidence for a correlation between the stellar population age and \pwind, with older stellar populations having lower pressures for the same radius and mass. 

We expect that the radius and luminosity of the \HII\ regions are correlated as $r$\,$\propto$\,$Q^{1/3}$\,$\propto$\,\Lbol$^{1/3}$ for a constant \ne\ (e.g. \citealp{stromgren_1939}). 
This could complicate our interpretation of how the \HII\ regions evolve through the $r-P$ parameter space; 
i.e., for fixed age and \ne, a more luminous stellar population will produce a larger \HII\ region. 
Normalising $r$ by $\Lbol^{1/3}$ accounts for differences in ionising photon production between regions without assuming a single representative value of \ne, which our observations show varies from region to region.
We see a tight correlation between $r/\Lbol^{1/3}$ and $P$ for all terms (Fig\,\ref{fig:PvRL}).
Moreover, we find weak trends that for a given $r/\Lbol^{1/3}$ and mass, younger \HII\ regions appear to have systematically higher $P$ (again this is most evident for \pwind).

\begin{figure}
    \centering
	\includegraphics[width=\columnwidth]{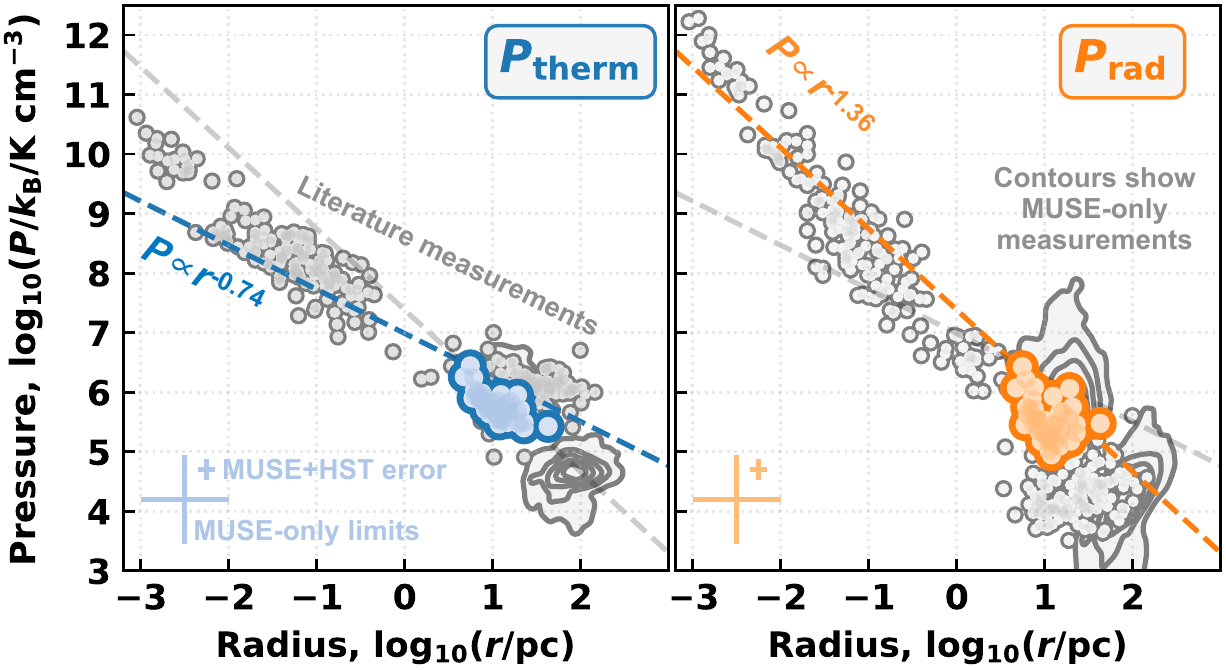}
    \caption{{\bf The thermal {\it (left panel)} and direct radiation {\it (right panel)} pressure terms.} Coloured points are \HII\ regions studied in this work, grey points are measurements from the literature \citep{lopez_2014,mcleod_2019, mcleod_20, McLeod2021, barnes20b, olivier2020}, and the contours show the upper and lower limits for all galaxies studied by \citet{Barnes2021b}. 
    The dashed lines show the power law relations from \citet{olivier2020}.}
    \label{fig:PvRlit}
\end{figure}

We also investigate how the local environment affects the various feedback mechanisms. 
Fig.\,\ref{fig:PvRmetal} shows the pressures as a function of the galactocentric radius ($R_\mathrm{gal}$). We see a weak trend of increased pressure at higher $R_\mathrm{gal}$. 
These higher pressures correlate with a weak decrease in the metallicity, or an increase in extinction. 
The increase of \ptherm\ and \pdir\ with $R_\mathrm{gal}$ could be due to higher photon fluxes from lower-metallicity stellar populations (c.f. \citealp{McLeod2021}). 
The increase in \pwind\ at lower metallicity is, however, not expected (e.g. \citealp{Kudritzki2002}), but more data is needed to further explore this trend. 

We compare the pressures determined in this work to those estimated in the literature (Fig.\,\ref{fig:PvRlit}). 
Here we compare only the thermal (shown in blue) and radiation (shown in orange) pressure terms, which are typically calculated in a similar manner as in this work.
We see that the pressures determined here are consistent with those in the literature. 
In Fig.\,\ref{fig:PvRlit} we also indicate the size-pressure relations from \citet{olivier2020}, for which the $P_\mathrm{rad}\,\propto\,R^{-0.74}$ and $P_\mathrm{therm}\,\propto\,R^{-1.36}$ appear to be in good agreement with both our literature compilation and measured values.\footnote{Note that the range of bolometric luminosities of the \HII\ regions studied in this work ($10^{5.5-7.3}$\,\Lsol) are at the upper end of the range from the \citet{olivier2020} sample ($10^{3.5-6.4}$\,\Lsol).}
These relations cross at sizes similar to our smallest \HII\ region ($\sim5\,pc$). 
Therefore, we are probing an interesting part of parameter space, where the single scattering radiation pressure and thermal pressure of ionized gas make comparable (to within a factor of a few) contributions to the expansion of \HII\ regions.


In the lower-left of the panels of Fig.\,\ref{fig:PvRlit} we also show a representative uncertainty range spanned by the limits calculated just using ground-based observations \citep{Barnes2021b}, which can be compared to the error range of the estimates from this work. 
The order of magnitude reduction error is primarily due to the inclusion of the resolved size measurements from the HST \Ha\ maps, which all the pressure terms depend on strongly. 
An additional gain is also achieved by the inclusion of the stellar population information, which is included within all the pressure terms. 
Specifically, the stellar population age associated with each \HII\ region is used (with the extinction corrected \Ha\ luminosity from MUSE) to determine its bolometric luminosity that is used in the calculation of \ptherm\ and \pdir, and both the stellar population age and mass are to infer the mechanical luminosity and mass loss rate that is used in the calculation of \pwind.

In all, we have demonstrated in this letter that, together, HST and MUSE significantly improved our constraints on the feedback pressures dominating within \HII\ regions across nearby galaxies (c.f. Fig.\,\ref{fig:PvR}). This represents an important step in directly connecting how young stars destroy their host environments, which can in future be expanded to larger samples of \HII\ regions and more galactic environments, and provide a statistical benchmark for the inclusion of pre-SNe stellar feedback in theory and simulations; e.g. from resolved planet, disc and/or star formation simulations, to large-scale galaxy dynamic and evolution simulations, and cosmological simulations.

\begin{acknowledgements}
    We would like to thank the referee for their constructive feedback that helped improve the quality of this paper.
    This work was carried out as part of the PHANGS collaboration, based on observations collected at the European Southern Observatory under ESO programmes 1100.B-0651 (PHANGS--MUSE; PI Schinnerer). 
    ATB and FB would like to acknowledge funding from the European Research Council (ERC) under the European Union’s Horizon 2020 research and innovation programme (grant agreement No.726384/Empire).
    MB gratefully acknowledges support by the ANID BASAL project FB210003 and from the FONDECYT regular grants 1211000.
    RSK \& SCOG acknowledge funding from the Deutsche Forschungsgemeinschaft (DFG) via SFB 881 `The Milky Way System' (subprojects A1, B1, B2 and B8) and from the Heidelberg Cluster of Excellence {\em STRUCTURES} in the framework of Germany’s Excellence Strategy (grant EXC-2181/1 - 390900948). They also acknowledge support from the European Research Council in the ERC synergy grant `ECOGAL – Understanding our Galactic ecosystem: From the disk of the Milky Way to the formation sites of stars and planets' (project ID 855130).
    HAP acknowledges support by the Ministry of Science and Technology of Taiwan under grant 110-2112-M-032-020-MY3.
    EC acknowledges support from ANID project Basal AFB-170002.
    MC and JMDK gratefully acknowledge funding from the German Research Foundation (DFG) in the form of an Emmy Noether Research Group (grant number KR4801/1-1), as well as from the European Research Council (ERC) under the European Union's Horizon 2020 research and innovation programme via the ERC Starting Grant MUSTANG (grant agreement number 714907).
    KK, FS and OE gratefully acknowledge funding from the German Research Foundation (DFG) in the form of an Emmy Noether Research Group (grant number KR4598/2-1, PI Kreckel).
    ES and TGW acknowledge funding from the European Research Council (ERC) under the European Union’s Horizon 2020 research and innovation programme (grant agreement No. 694343).
\end{acknowledgements}

\bibliographystyle{aa}
\bibliography{references}

\begin{appendix}
\section{Sample properties and data}

In table\,\ref{tab1}, we summarise the properties of the \HII\ region sample studied in this work. A machine-readable version of this table is available online. 

Science-level MUSE mosaicked datacubes and high-level analysis products (e.g. emission line fluxes) are provided via the ESO archive phase 3 interface.\footnote{\url{https://archive.eso.org/scienceportal/home?data_collection=PHANGS}} A full description of the first PHANGS-MUSE data release is presented in \citet{Emsellem2022}. Science-level HST broadband images and higher-level data products are available in the MAST archive,\footnote{\url{https://archive.stsci.edu/hlsp/phangs-hst}}. A full description of the first PHANGS-HST data release is presented in \citet{Lee2021}. The HST \Ha\ images are publicly available in the Hubble Legacy Archive.\footnote{\url{https://hla.stsci.edu/}}

\begin{sidewaystable*}
\caption{{\bf Properties of the \HII\ region sample.} We show in columns from left to right the ID, central RA and Dec, radius, galactocentric radius (normalised by the effective radius), extinction corrected \Ha\ flux, extinction, extinction corrected \Ha\ luminosity, metallicity, stellar population age and mass, bolometric and mechanical luminosity, mass loss rate, electron density, and thermal, radiation and wind pressures. A machine-readable version of this table is available online.}\label{tab1}
\centering
\begin{tabular}{ccccccccccccccccccc}
\hline\hline
ID & RA & Dec & $r$ & $R_\mathrm{gal}/R_\mathrm{eff}$ & $F_\mathrm{H\alpha,corr}$ & EBV & $L_\mathrm{H\alpha}$ & metal. & $t$ & $M$ & $L_\mathrm{bol}$ & $L_\mathrm{mech}$ & $\dot M$ & $n_\mathrm{e}$ & \ptherm & \pdir & \pwind \\
 & & & & & log$_{10}$() & & log$_{10}$() & & & log$_{10}$() & log$_{10}$() & log$_{10}$() & log$_{10}$() & & log$_{10}$() & log$_{10}$() & log$_{10}$() \\
 & $\mathrm{{}^{\circ}}$ & $\mathrm{{}^{\circ}}$ & $\mathrm{pc}$ &  & $\mathrm{erg\,s^{-1}\,cm^{-2}}$ & $\mathrm{mag}$ & $\mathrm{L_{\odot}}$ & 12+log(O/H) & $\mathrm{Myr}$ & $\mathrm{M_{\odot}}$ & $\mathrm{L_{\odot}}$ & $\mathrm{erg\,s^{-1}}$ & $\mathrm{g\,s^{-1}}$ & $\mathrm{cm^{-3}}$ & $\mathrm{K\,cm^{-3}}$ & $\mathrm{K\,cm^{-3}}$ & $\mathrm{K\,cm^{-3}}$ \\
 \hline
0 & 71.452 & -59.256 & 10.1 & 1.6 & -14.95 & 0.144 & 4.12 & 8.58 & 1 & 3.71 & 5.9 & 37.7 & 20.9 & 30.4 & 5.69 & 5.26 & 5.65 \\
1 & 71.400 & -59.260 & 12.9 & 2.58 & -14.93 & 0.277 & 4.14 & 8.56 & 2 & 3.49 & 5.99 & 37.4 & 20.6 & 21.8 & 5.54 & 5.14 & 5.23 \\
2 & 71.389 & -59.249 & 42.6 & 2.36 & -13.48 & 0.301 & 5.58 & 8.55 & 1 & 5 & 7.37 & 38.9 & 22.1 & 18.3 & 5.43 & 5.47 & 5.69 \\
3 & 71.476 & -59.232 & 20.6 & 3.94 & -13.84 & 0.248 & 5.23 & 8.56 & 1 & 4.3 & 7.01 & 38.2 & 21.4 & 36 & 5.71 & 5.75 & 5.62 \\
4 & 71.459 & -59.246 & 14.6 & 1.95 & -14.86 & 0.17 & 4.21 & 8.57 & 2 & 3.54 & 6.06 & 37.5 & 20.7 & 19.4 & 5.49 & 5.09 & 5.16 \\
5 & 71.437 & -59.245 & 22.7 & 0.752 & -14.47 & 0.3 & 4.59 & 8.61 & 4 & 3.52 & 6.8 & 37.5 & 20.7 & 15.7 & 5.4 & 5.46 & 4.75 \\
6 & 71.468 & -59.240 & 7.17 & 2.82 & -15.04 & 0.369 & 4.02 & 8.54 & 4 & 3.68 & 6.23 & 37.6 & 20.8 & 45.8 & 5.86 & 5.88 & 5.92 \\
7 & 71.483 & -59.234 & 15.3 & 4.13 & -14.46 & 0.185 & 4.61 & 8.52 & 1 & 3.93 & 6.4 & 37.9 & 21.1 & 28.8 & 5.66 & 5.39 & 5.51 \\
8 & 71.471 & -59.228 & 19.2 & 3.97 & -14.09 & 0.288 & 4.97 & 8.62 & 1 & 4.06 & 6.76 & 38 & 21.2 & 30.7 & 5.67 & 5.56 & 5.44 \\
9 & 71.428 & -59.236 & 9.99 & 1.42 & -14.69 & 0.22 & 4.38 & 8.58 & 1 & 3.8 & 6.16 & 37.7 & 20.9 & 41.9 & 5.83 & 5.53 & 5.76 \\
10 & 71.483 & -59.246 & 6.09 & 3.47 & -14.50 & 0.418 & 4.56 & 8.51 & 2 & 4.32 & 6.42 & 38.3 & 21.5 & 109 & 6.24 & 6.21 & 6.7 \\
11 & 71.380 & -59.250 & 4.74 & 2.93 & -14.79 & 0.358 & 4.27 & 8.58 & 1 & 3.65 & 6.06 & 37.6 & 20.8 & 114 & 6.26 & 6.07 & 6.25 \\
12 & 71.380 & -59.245 & 5.56 & 2.77 & -14.36 & 0.278 & 4.70 & 8.52 & 2 & 3.91 & 6.55 & 37.8 & 21.1 & 154 & 6.44 & 6.43 & 6.37 \\
13 & 71.376 & -59.240 & 11.3 & 2.96 & -14.90 & 0.296 & 4.17 & 8.58 & 1 & 3.95 & 5.95 & 37.9 & 21.1 & 31.8 & 5.86 & 5.21 & 5.8 \\
14 & 71.414 & -59.235 & 14.7 & 1.45 & -14.22 & 0.563 & 4.84 & 8.61 & 1 & 4.43 & 6.63 & 38.4 & 21.6 & 40.1 & 5.81 & 5.66 & 6.05 \\
15 & 71.374 & -59.251 & 8.26 & 3.36 & -14.83 & 0.238 & 4.24 & 8.54 & 1 & 3.64 & 6.02 & 37.6 & 20.8 & 47.4 & 5.88 & 5.55 & 5.75 \\
16 & 71.369 & -59.257 & 8.63 & 4.08 & -15.08 & 0.125 & 3.99 & 8.52 & 2 & 3.43 & 5.84 & 37.4 & 20.6 & 33.5 & 5.73 & 5.34 & 5.51 \\
17 & 71.385 & -59.241 & 12.5 & 2.45 & -14.09 & 0.35 & 4.97 & 8.54 & 1 & 4.17 & 6.76 & 38.1 & 21.3 & 58.8 & 5.96 & 5.93 & 5.93 \\
18 & 71.392 & -59.237 & 11.8 & 2.11 & -14.78 & 0.171 & 4.28 & 8.55 & 1 & 3.94 & 6.07 & 37.9 & 21.1 & 29.1 & 5.67 & 5.29 & 5.75 \\
19 & 71.375 & -59.250 & 19.2 & 3.3 & -13.66 & 0.308 & 5.41 & 8.51 & 2 & 4.17 & 7.26 & 38.1 & 21.3 & 53 & 5.95 & 6.06 & 5.55 \\
20 & 71.451 & -59.259 & 11.3 & 1.67 & -14.95 & 0.239 & 4.12 & 8.54 & 1 & 3.42 & 5.9 & 37.4 & 20.6 & 25.9 & 5.62 & 5.16 & 5.27 \\
21 & 71.480 & -59.243 & 7.41 & 3.4 & -15.03 & 0.205 & 4.04 & 8.50 & 1 & 3.68 & 5.82 & 37.6 & 20.8 & 44.5 & 5.85 & 5.45 & 5.89 \\
22 & 71.402 & -59.244 & 7.58 & 1.48 & -14.74 & 0.327 & 4.33 & 8.56 & 1 & 3.99 & 6.11 & 37.9 & 21.1 & 59.9 & 5.98 & 5.72 & 6.18 \\
23 & 71.379 & -59.243 & 5.97 & 2.81 & -15.21 & 0.233 & 3.86 & 8.54 & 1 & 3.37 & 5.64 & 37.3 & 20.5 & 49.8 & 5.9 & 5.45 & 5.77 \\
24 & 71.391 & -59.244 & 22.8 & 2.13 & -14.25 & 0.106 & 4.81 & 8.54 & 2 & 3.46 & 6.66 & 37.4 & 20.6 & 19.9 & 5.49 & 5.32 & 4.7 \\
25 & 71.421 & -59.241 & 9.12 & 0.727 & -15.18 & 0.259 & 3.88 & 8.59 & 2 & 3.3 & 5.73 & 37.2 & 20.4 & 27.1 & 5.64 & 5.18 & 5.32 \\
26 & 71.433 & -59.237 & 13.4 & 1.42 & -14.49 & 0.257 & 4.57 & 8.58 & 1 & 4.06 & 6.36 & 38 & 21.2 & 34 & 5.74 & 5.47 & 5.76 \\
27 & 71.395 & -59.240 & 11.2 & 1.89 & -15.10 & 0.144 & 3.97 & 8.56 & 1 & 3.75 & 5.76 & 37.7 & 20.9 & 22.1 & 5.55 & 5.02 & 5.6 \\
28 & 71.403 & -59.260 & 8.27 & 2.37 & -15.18 & 0.305 & 3.88 & 8.56 & 3 & 3.45 & 6.01 & 37.4 & 20.6 & 31.5 & 5.7 & 5.54 & 5.57 \\
29 & 71.454 & -59.256 & 8.07 & 1.68 & -14.96 & 0.131 & 4.11 & 8.54 & 2 & 3.6 & 5.96 & 37.5 & 20.7 & 42.4 & 5.83 & 5.51 & 5.74 \\
30 & 71.478 & -59.244 & 6.43 & 3.2 & -15.05 & 0.103 & 4.02 & 8.51 & 2 & 3.39 & 5.87 & 37.3 & 20.5 & 53.6 & 5.93 & 5.62 & 5.73 \\
31 & 71.469 & -59.227 & 6.22 & 3.98 & -14.88 & 0.264 & 4.19 & 8.59 & 1 & 3.93 & 5.97 & 37.9 & 21.1 & 68.4 & 6.04 & 5.75 & 6.29 \\
32 & 71.453 & -59.242 & 11.9 & 1.9 & -15.18 & 0.241 & 3.88 & 8.59 & 2 & 2.89 & 5.73 & 36.8 & 20 & 18.3 & 5.47 & 4.95 & 4.69 \\
33 & 71.418 & -59.256 & 12.5 & 1.36 & -15.01 & 0.194 & 4.05 & 8.59 & 4 & 3.03 & 6.26 & 37 & 20.2 & 20.5 & 5.52 & 5.43 & 4.78 \\
34 & 71.418 & -59.257 & 15 & 1.42 & -14.58 & 0.31 & 4.49 & 8.63 & 1 & 3.8 & 6.27 & 37.7 & 20.9 & 27 & 5.68 & 5.29 & 5.4 \\
35 & 71.408 & -59.261 & 19.7 & 2.23 & -14.52 & 0.249 & 4.55 & 8.55 & 2 & 3.7 & 6.4 & 37.6 & 20.8 & 18.5 & 5.47 & 5.18 & 5.06 \\
36 & 71.396 & -59.248 & 6.69 & 1.93 & -14.85 & 0.337 & 4.22 & 8.60 & 1 & 3.68 & 6 & 37.6 & 20.8 & 63.5 & 6.01 & 5.72 & 5.98 \\
37 & 71.416 & -59.255 & 15.9 & 1.29 & -14.15 & 0.387 & 4.92 & 8.60 & 1 & 4.32 & 6.7 & 38.3 & 21.5 & 40.3 & 5.85 & 5.67 & 5.87 \\
38 & 71.463 & -59.244 & 7.62 & 2.28 & -14.79 & 0.39 & 4.27 & 8.56 & 4 & 3.65 & 6.48 & 37.6 & 20.8 & 55.8 & 5.95 & 6.08 & 5.84 \\
39 & 71.451 & -59.251 & 8.59 & 1.38 & -15.03 & 0.166 & 4.03 & 8.59 & 2 & 3.32 & 5.88 & 37.3 & 20.5 & 35.4 & 5.75 & 5.38 & 5.4 \\
\hline\hline
\end{tabular}
\end{sidewaystable*}

\end{appendix}

\end{document}